\documentclass[fleqn,usenatbib]{mnras}

\usepackage{newtxtext,newtxmath}

\usepackage[T1]{fontenc}

\DeclareRobustCommand{\VAN}[3]{#2}
\let\VANthebibliography\thebibliography
\def\thebibliography{\DeclareRobustCommand{\VAN}[3]{##3}\VANthebibliography}

\usepackage{graphicx}	
\usepackage{amsmath}	
\usepackage{subcaption}
\defcitealias{Meneghetti_et_al_2020}{M20}
\newcommand{\msun}{\mathrm{M}_\odot}
\newcommand{\mvir}{M_\mathrm{vir}}
\newcommand{\rvir}{r_\mathrm{vir}}
\newcommand{\vmax}{v_\mathrm{max}}
\newcommand{\msub}{M_\mathrm{sub}}

\title[Substructures in simulated clusters]{Strongly lensed cluster substructures are not in tension with $\Lambda$CDM}

\author[Y. M. Bah\'{e}]{
Yannick M. Bah\'{e}$^{1}$\thanks{E-mail: bahe@strw.leidenuniv.nl}
\\
$^{1}$Leiden Observatory, Leiden University, PO Box 9513, 2300 RA Leiden, The Netherlands\\
}

\date{Accepted XXX. Received YYY; in original form ZZZ}

\pubyear{2021}

\begin{document}
\label{firstpage}
\pagerange{\pageref{firstpage}--\pageref{lastpage}}
\maketitle

\begin{abstract}
Strong gravitational lensing observations can test structure formation models by constraining the masses and concentrations of subhaloes in massive galaxy clusters. Recent work has concluded that cluster subhaloes are more abundant and/or concentrated than predicted by $\Lambda$CDM simulations; this finding has been interpreted as arising from unidentified issues with  simulations or an incorrect understanding of the nature of dark matter. We test these hypotheses by comparing observed subhalo masses and maximum circular velocities $\vmax$ to predictions from the high resolution Hydrangea galaxy cluster simulation suite, which is based on the successful EAGLE galaxy formation model. The simulated subhalo mass distribution and mass--$\vmax$ relation agrees well with observations, due to the presence of baryons during tidal stripping. Similar agreement is found for the lower-resolution Illustris-TNG300 simulation. In combination, our results suggest that the abundance and concentration of cluster substructures are not in tension with $\Lambda$CDM, but may provide useful constraints for the refinement of baryon physics models in simulations.
\end{abstract}




\section{Introduction}

In a $\Lambda$ Cold Dark Matter ($\Lambda$CDM) universe, structures from dwarf galaxies to massive galaxy clusters have formed hierarchically, through mergers and accretion of smaller structures \citep{Blumenthal_et_al_1984}. There is a plethora of observational evidence for this general picture, including the ubiquity of stellar haloes and tidal streams within galaxies (e.g.~\citealt{Helmi_2008,Shipp_et_al_2018}) and -- perhaps most directly -- the existence of satellite galaxies and their associated dark matter ``subhaloes'' that have not yet merged (completely) with their host halo (e.g.~\citealt{Yang_et_al_2007,Grillo_et_al_2015}).

Massive substructures near the centre of massive galaxy clusters represent a particularly attractive quantitative test of $\Lambda$CDM: their gravitational potential combined with that of their host cluster gives rise to strong lensing distortions of background galaxies, from which their mass distribution can be reconstructed in an, in principle, direct and unbiased way (e.g.~\citealt{Kneib_Natarajan_2011}). The lens models derived from these observations typically prefer a larger number of massive cluster substructures than predicted by $\Lambda$CDM $N$-body simulations (e.g.~\citealt{Grillo_et_al_2015}).

While this could be interpreted as a failure of the $\Lambda$CDM paradigm, a more trivial solution is that the presence of baryons modifies the properties of subhaloes in ways not captured by $N$-body simulations. Feedback from star formation or active galactic nuclei, for example, may expel gas from galaxies and thus lower their central concentration, while gas condensation and star formation have the opposite effect (e.g.~\citealt{Duffy_et_al_2010,Despali_Vegetti_2017}). A meaningful test of the $\Lambda$CDM paradigm therefore requires comparisons to hydrodynamical simulations that self-consistently include these baryonic processes. Unlike their numerically well-converged $N$-body analogues (e.g.~\citealt{Wang_et_al_2020}), the development and refinement of such simulations is still an ongoing effort (see e.g.~\citealt{Vogelsberger_et_al_2020}), and their considerably higher computational cost typically restricts hydrodynamic simulations of galaxy clusters to much lower resolution than equivalent $N$-body calculations (e.g.~\citealt{Gao_et_al_2012,Cui_et_al_2018}).

In a recent study, \citet[][hereafter M20]{Meneghetti_et_al_2020} compared substructure detections in a sample of eleven intermediate-redshift ($z \approx 0.4$) clusters to a suite of cosmological hydrodynamic simulations \citep{Planelles_et_al_2014} and still found significant tensions: the observed clusters produce an order of magnitude more strong lensing events than simulated clusters, and the inferred maximum circular velocity of cluster subhaloes -- a proxy for their concentration -- is several times larger than predicted. In view of the modelling uncertainties and limited resolution of such simulations, the interpretation of this discrepancy requires comparisons to additional simulations with different resolution and modelling approaches. \citetalias{Meneghetti_et_al_2020} include some comparisons of this nature, none of which fully resolve the tension: they therefore interpret their findings as evidence of either a so-far unappreciated failure of cosmological simulations, or the need to consider alternatives to $\Lambda$CDM (see e.g.~\citealt{Yang_Yu_2021}).

We extend this approach by comparing their observations to galaxy clusters from the state-of-the-art cosmological hydrodynamical Hydrangea/C-EAGLE simulation suite. They achieve a mass resolution $\sim$100 times higher than the \citet{Planelles_et_al_2014} simulations used by \citetalias[][]{Meneghetti_et_al_2020}, and the underlying EAGLE baryon physics model \citep{Schaye_et_al_2015} has been shown to reproduce a wide variety of observations on galaxy and cluster scales (see e.g. \citealt{Schaye_et_al_2015,McAlpine_et_al_2016,Bahe_et_al_2017}, and references therein). As such, they represent the arguably best tool to confront the observations of \citetalias[][]{Meneghetti_et_al_2020} with expectations from the $\Lambda$CDM model.

The remainder of this letter is structured as follows. In Section \ref{sec:sim_obs}, we summarize the key aspects of the simulations and of the observational analysis of \citetalias[][]{Meneghetti_et_al_2020}. We compare the two in Section \ref{sec:comparison}; surprisingly, we find good agreement. In Section \ref{sec:baryons}, we test the importance of baryon physics for our results, before summarising our results in Section \ref{sec:summary}.  

\section{Overview of observations and simulations}
\label{sec:sim_obs}

\subsection{Strong lensing observations}
\citetalias{Meneghetti_et_al_2020} characterize the substructure population in 11 massive galaxy clusters in the redshift range $0.2 < z < 0.6$  (median $z = 0.39$) from the Hubble Frontier Fields \citep{Lotz_et_al_2017} and CLASH \citep{Postman_et_al_2012} surveys. \textit{Hubble Space Telescope} lensing observations from these surveys, combined with spectroscopic data from the ESO \textit{Very Large Telescope} for a subset of clusters \citep{Treu_et_al_2015,Caminha_et_al_2016}, were modelled with the \textsc{Lenstool} code \citep{Jullo_et_al_2007}. For their main ``reference'' sample of three clusters, on which we focus here, this model consists of a combination of cluster-scale and galaxy-scale dual pseudo-isothermal elliptical (dPIE) mass distributions; the latter are centred on the locations of galaxies and their two free parameters are constrained by power-law scaling relations with the galaxy luminosity that themselves incorporate measurements of stellar kinematics \citep{Bergamini_et_al_2019}. Subhalo masses are then obtained from integrating each dPIE profile, while their maximum circular velocities $\vmax \equiv \sqrt{GM(< r) / r}$ (where $G$ is Newton's constant, and $M(< r)$ the mass contained within radius $r$) are directly related to the dPIE parameter $\sigma_0$ by $\vmax = \sqrt{2}\sigma_0$. For further details, the reader is referred to \citetalias{Meneghetti_et_al_2020}.

\subsection{The Hydrangea simulations}
\subsubsection{Simulation overview}
The Hydrangea simulations \citep{Bahe_et_al_2017, Barnes_et_al_2017} are a suite of 24 simulated massive galaxy clusters ($10^{14} < M_\mathrm{200c} < 2.5 \times 10^{15}$ at $z = 0$) and their large-scale surroundings\footnote{$M_\mathrm{200c}$ is the mass within a radius $r_\mathrm{200c}$ from the cluster centre of potential, within which the mean density equals 200 times the critical density of the Universe at that redshift; $\mvir$ and $\rvir$ denote the corresponding values at an overdensity $\Delta_\mathrm{c} (z)$ corresponding to a collapsed spherical top-hat perturbation \citep{Bryan_Norman_1998}. For consistency with \citetalias{Meneghetti_et_al_2020}, we mostly use the latter definition in this work.}. They are part of the C-EAGLE project of zoom-in cluster simulations with a variant of the EAGLE simulation code \citep{Schaye_et_al_2015} and have a mass resolution of $\approx$1.8 $\times 10^6\, \msun$ for baryons and $\approx$9.7 $\times 10^6\, \msun$ for dark matter, respectively; the gravitational softening is 700 pc at $z < 2.8$. The simulation code is based on the \textsc{anarchy} variant of Smoothed Particle Hydrodynamics \citep{Schaller_et_al_2015} and incudes sub-grid models for the same astrophysical processes as the simulations of \citet{Planelles_et_al_2014}, but often implemented in substantially different ways: element-by-element radiative cooling and photoheating \citep{Wiersma_et_al_2009a}, reionization \citep{Wiersma_et_al_2009b}, star formation \citep[][with the metallicity-dependent star formation threshold of \citealt{Schaye_2004}]{Schaye_DallaVecchia_2008}, mass and metal enrichment from stellar outflows \citep{Wiersma_et_al_2009b}, energy feedback from star formation in stochastic thermal form \citep{DallaVecchia_Schaye_2012}, and for the seeding, growth, and energy feedback from supermassive black holes \citep{Rosas-Guevara_et_al_2015, Schaye_et_al_2015}. The implementation of these models is described in detail by \citet{Schaye_et_al_2015} and \citet{Bahe_et_al_2017}.

One key aspect of the EAGLE model -- and together with the higher resolution the main difference from the \citet{Planelles_et_al_2014} simulations analysed by \citetalias{Meneghetti_et_al_2020} -- is that sub-grid model parameters that are not well-constrained by observations (specifically, the scaling of energy feedback from star formation with local properties and the coupling efficiency of AGN feedback) were calibrated against the stellar mass function, galaxy sizes, and black hole masses of field galaxies in the local Universe \citep{Crain_et_al_2015}. No cluster-scale observations were considered in the calibration process; as a consequence, some properties of the cluster haloes are in tension with observations, in particular the total gas fractions \citep{Barnes_et_al_2017} and iron abundance \citep{Pearce_et_al_2020} of the intra-cluster medium, and the stellar masses of the central cluster galaxies \citep{Bahe_et_al_2017}. The stellar mass function of satellite galaxies, on the other hand, matches observations closely both at $z = 0$ \citep{Bahe_et_al_2017} and $z \approx 0.6$ \citep{Ahad_et_al_2021}.

\subsubsection{Subhalo identification}
\label{sec:subhaloes}
Subhaloes are found in the simulation outputs with the \textsc{subfind} algorithm \citep[see also \citealt{Springel_et_al_2001}]{Dolag_et_al_2009}, which identifies gravitationally self-bound particles within locally overdense regions of  friends-of-friends haloes. This algorithm has a known tendency to miss some particles that are physically part of a subhalo, and hence underestimate their true mass, in the central regions of rich clusters (e.g.~\citealt{Muldrew_et_al_2011,Behroozi_et_al_2013}). To avoid this, we use the re-computed ``Cantor'' subhalo catalogue as described in Bah\'{e} et al. (in prep.): in brief, this method is based on the \textsc{subfind} catalogues, but also considers particles that belonged to the progenitor of a subhalo in previous snapshots. Bah\'{e} et al. (in prep.) show that, in contrast to \textsc{subfind}, this approach leads to an almost perfect separation between subhaloes and the smooth cluster halo; subhalo masses are therefore systematically higher than in \textsc{subfind} by factors of $\approx$1.25--10 (highest for the most massive subhaloes and closest to the cluster centre), but maximum circular velocities ($\vmax$) typically differ by only a few per cent.

A caveat is that the subhalo properties from Cantor (and also \textsc{subfind}) are based on physical considerations, rather than mimicking the observational procedure. With our primary aim being to test the presence of a large, fundamental offset between the simulations and observations, we neglect the impact of this difference here and instead compare simulations and observations at face value.

\section{The accuracy of simulated cluster substructures}
\label{sec:comparison}

\subsection{Predicted subhalo mass functions}
\begin{figure}
	\includegraphics[width=\columnwidth]{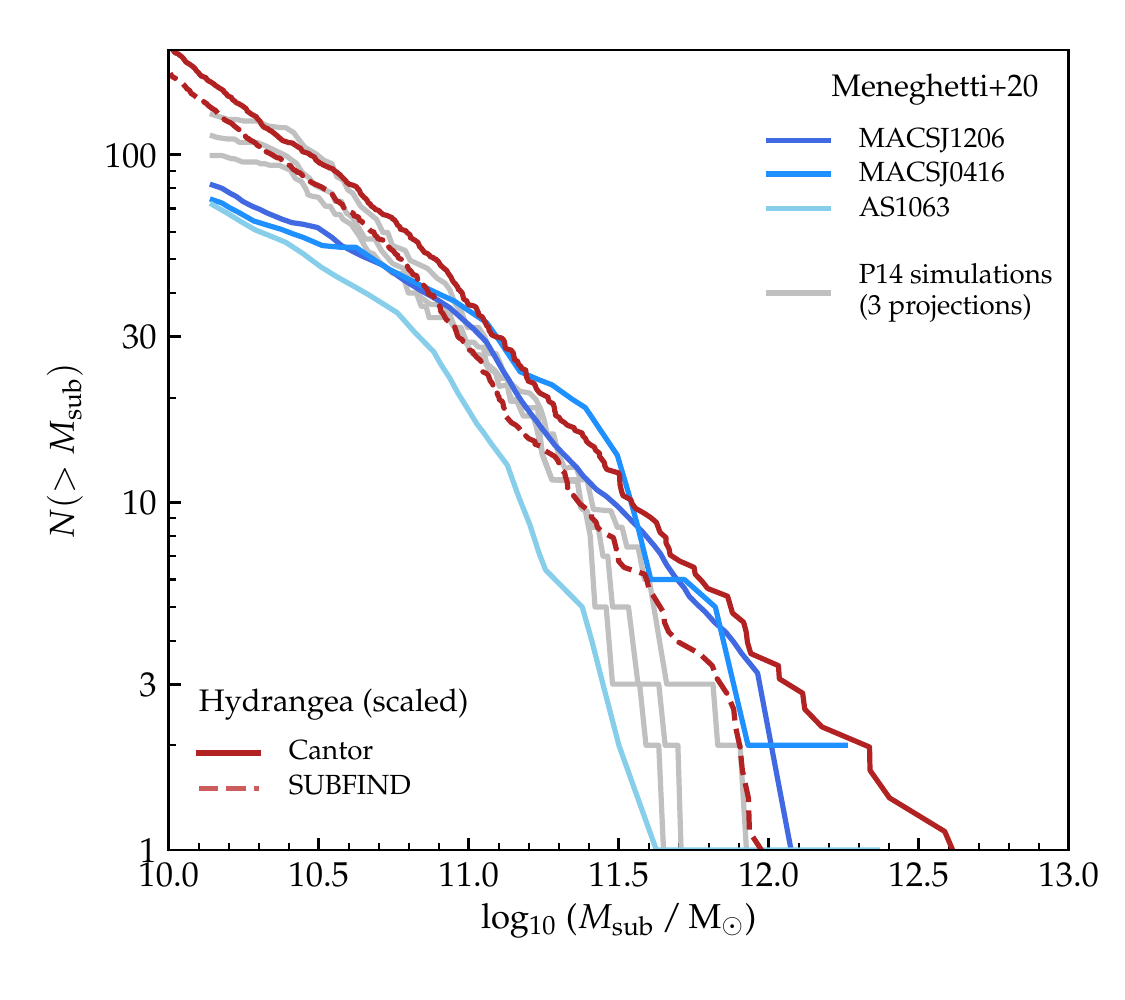}
	\caption{The cumulative subhalo mass function within $R_\mathrm{2D} < 0.15 \rvir$ as predicted by Hydrangea (red) and measured observationally (blue shades, different lines represent the three different `Reference' clusters of \citetalias{Meneghetti_et_al_2020}). For Hydrangea, solid (dashed) lines represent subhaloes identified with the Cantor (\textsc{subfind}) structure finders, respectively, and subhalo counts are scaled to account for the offset in cluster mass with respect to the observations (see text). The three grey lines represent the subhalo mass function in the simulations of \citet{Planelles_et_al_2014} in three different projections. Both simulations predict subhalo mass functions in good agreement with the observations, at least at $M_\text{sub} \gtrsim 3 \times 10^{10}\, \msun$.}
	\label{fig:mass_function}
\end{figure}

We begin by comparing the (cumulative) subhalo mass function predicted by Hydrangea to the lensing observations in Fig.~\ref{fig:mass_function}. From the simulation snapshots at redshift $z = 0.41$, we select 9 clusters with $\mvir > 5 \times 10^{14}\, \msun$ whose centre is at least 8 co-moving Mpc away from any low-resolution boundary particles. In each cluster, we then select subhaloes with mass $M_\mathrm{sub} > 10^{10}\, \msun$, projected distance (in the simulation \emph{xy} plane) from the cluster centre of potential of $R_\mathrm{2D} \leq 0.15\,\rvir$, and with a maximum offset of $2\times \rvir$ along the $z$ direction. Finally, we exclude a small number of subhaloes very close to the cluster centres ($r_\mathrm{3D} < 10^{-3}\, r_\mathrm{vir}$) that are in the process of merging with the BCG and would not be detected as separate galaxies in observations. Our total sample contains 726 subhaloes.

The mean $M_\mathrm{200c}$ of these 9 Hydrangea clusters is $6.3 \times 10^{14}\, \msun$, which is noticeably lower than the three `reference' clusters analysed by \citetalias{Meneghetti_et_al_2020}: adopting masses of $M_\mathrm{200c} = 1.4 \times 10^{15}\, \msun$ \citep{Biviano_et_al_2013}, $9 \times 10^{14}\, \msun$ \citep{Balestra_et_al_2016}, and $2.5 \times 10^{15}\, \msun$ \citep{Sartoris_et_al_2020} for MACSJ1206, MACSJ0416, and AS1063, respectively, gives a mean of $1.6 \times 10^{15}\, \msun$. To account for this difference in halo mass, we scale the Hydrangea mass function\footnote{We have verified that all our results are qualitatively unchanged, and remain quantitatively consistent, when analysing only the most massive Hydrangea cluster ($M_\mathrm{200c} = 9.1 \times 10^{14}\,\msun$ at $z = 0.41$).} by a factor of 2.54.

The result, using subhalo masses from Cantor, is shown as the solid red line in Fig.~\ref{fig:mass_function}: the cumulative mass function is an almost straight power-law with an index close to 1, which agrees well with the observed distribution from \citetalias{Meneghetti_et_al_2020} (blue lines) down to $\msub \approx 3 \times 10^{10}\, \msun$. At lower masses, the observations show a noticeable flattening, plausibly due to detection incompleteness. The Hydrangea prediction also agrees reasonably well with the simulations of \citet{Planelles_et_al_2014} as shown by \citetalias{Meneghetti_et_al_2020} (grey lines), although these, like the observations, show a flattening at the low-mass end.

Not surprisingly, using the subhalo masses from \textsc{subfind} instead of Cantor (red dashed line in Fig.~\ref{fig:mass_function}) leads to a somewhat lower mass function, but only by a factor $\lesssim 2$. Neither of these methods matches the observational mass measurement in detail, but the relatively small difference between them indicates that the masses are not overly sensitive to subhalo finder details.

\subsection{Subhalo maximum circular velocities}
\begin{figure}
	\includegraphics[width=\columnwidth]{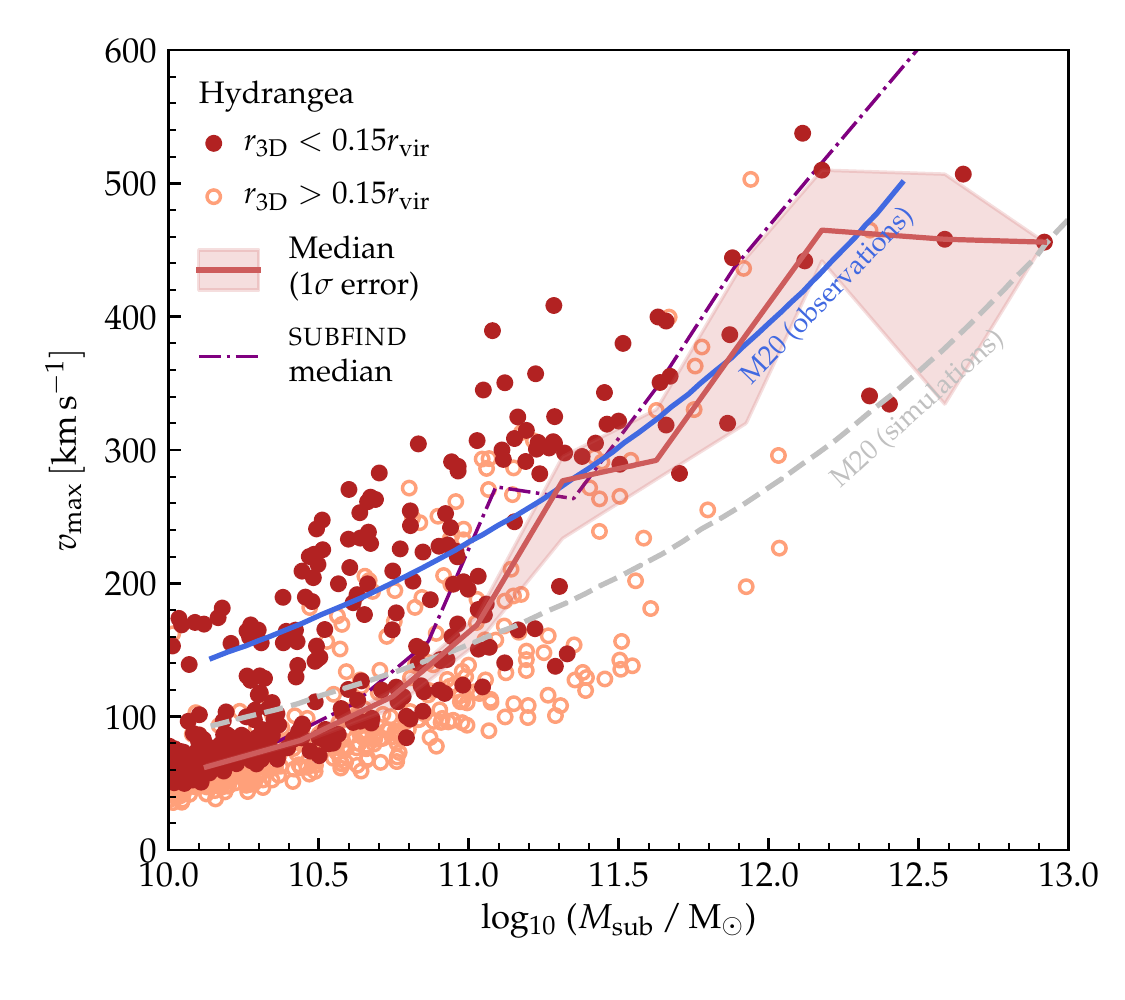}
	\caption{The relation between maximum circular subhalo velocity ($\vmax$) and subhalo mass. Red circles represent individual subhaloes in Hydrangea clusters at projected radii $R_\mathrm{2D} \leq 0.15 \rvir$; filled dark circles stand for subhaloes that are also physically close to the cluster centre ($r_\mathrm{3D} \leq 0.15\, \rvir$) whereas open light circles indicate subhaloes that are only close to the centre in projection. The running median for both samples combined and its $1\sigma$ uncertainty are shown, respectively, as a red solid line and light red shaded band; the purple dash-dotted line shows the corresponding median with subhalo properties from \textsc{subfind}. The blue solid and gray dashed lines give the relation inferred from the strong lensing observations of \citet{Meneghetti_et_al_2020} and their comparison simulations, respectively. At subhalo masses $\msub \gtrsim 3 \times 10^{10} \msun$, the Hydrangea simulations predict up to $\approx$2 times higher $\vmax$, in broad agreement with the observations.}
	\label{fig:vmax_msub}
\end{figure}

As \citetalias{Meneghetti_et_al_2020} have shown, broad agreement in the cumulative subhalo mass function is also achieved by the \citet{Planelles_et_al_2014} simulations. The strong difference in lensing signal instead stems from a discrepancy in the subhalo concentration, parameterised as the maximum circular velocity $\vmax$ (but see also 
\citealt{Robertson_2021} for the influence of resolution on the predicted lensing signal). We test the Hydrangea suite on this metric in Fig.~\ref{fig:vmax_msub}, using the same subhalo selection as above. We distinguish between subhaloes that are physically close to the cluster centre (3D distance $r_\mathrm{3D} < 0.15\,\rvir$, filled circles) and those that only appear close in projection (open circles); both should however be compared together to the observed relation. The (combined) running median in 0.25 dex bins in $\msub$ is shown as a red line with its 1$\sigma$ uncertainty from boostrapping indicated by the light red shaded region\footnote{We emphasize that this accounts only for the statistical uncertainty due to the limited Hydrangea sample size, not for the systematic error due to differences in the calculation of $\msub$ and $\vmax$ w.r.t. \citetalias{Meneghetti_et_al_2020}.}. As an indication of the sensitivity to subhalo definition, we also show the corresponding median trend using the \textsc{subfind} properties as a purple dash-dotted line.

The distribution of Hydrangea cluster subhaloes in the $M_\mathrm{sub}$--$\vmax$ plane is clearly bimodal: a lower sequence dominates at $M_\mathrm{sub} \lesssim 10^{11} \msun$ -- where we however caution that resolution effects may be non-negligible \citep{Schaye_et_al_2015} -- and roughly follows the trend from the \citet{Planelles_et_al_2014} simulations. The upper branch with $\approx$2--3 times higher $\vmax$ is more prominent at higher masses and \emph{follows the \citetalias{Meneghetti_et_al_2020} observational relation remarkably well}, as does the overall median at $\msub \gtrsim 2\times 10^{11}\,\msun$: if anything, these subhaloes have a slightly \emph{higher} $\vmax$ than measured\footnote{We note, however, that the observational relation is most strongly constrained by low-mass subhaloes (\citetalias{Meneghetti_et_al_2020}; Meneghetti, priv. comm.).}. The high-$\vmax$ branch is noticeably over-abundant in physically central subhaloes, which are also clustered closer to the top end within the branch\footnote{\citetalias{Meneghetti_et_al_2020} also show a $\msub$--$\vmax$ relation from Hydrangea, which falls well below their observations (their fig.~S10). The strong variation with cluster-centric radius resolves this apparent discrepancy: their relation is based on all galaxies within $\rvir$ (including those at large $r_\mathrm{2D}$), which are biased to larger radii than the galaxies probed in their observations.}. We will return to this point below.

\section{The role of baryon physics models}
\label{sec:baryons}
Compared to the \citet{Planelles_et_al_2014} simulations shown in \citetalias{Meneghetti_et_al_2020}, the Hydrangea suite differs in both resolution and the modelling of baryon physics. To gain insight into the origin of the higher $\vmax$ predicted by Hydrangea, we now test the role of the latter, by comparing the Hydrangea predictions first to an analogous suite of gravity-only runs and then to the Illustris-TNG300 simulation.

\subsection{The effect of baryons}
\begin{figure}
	\includegraphics[width=\columnwidth]{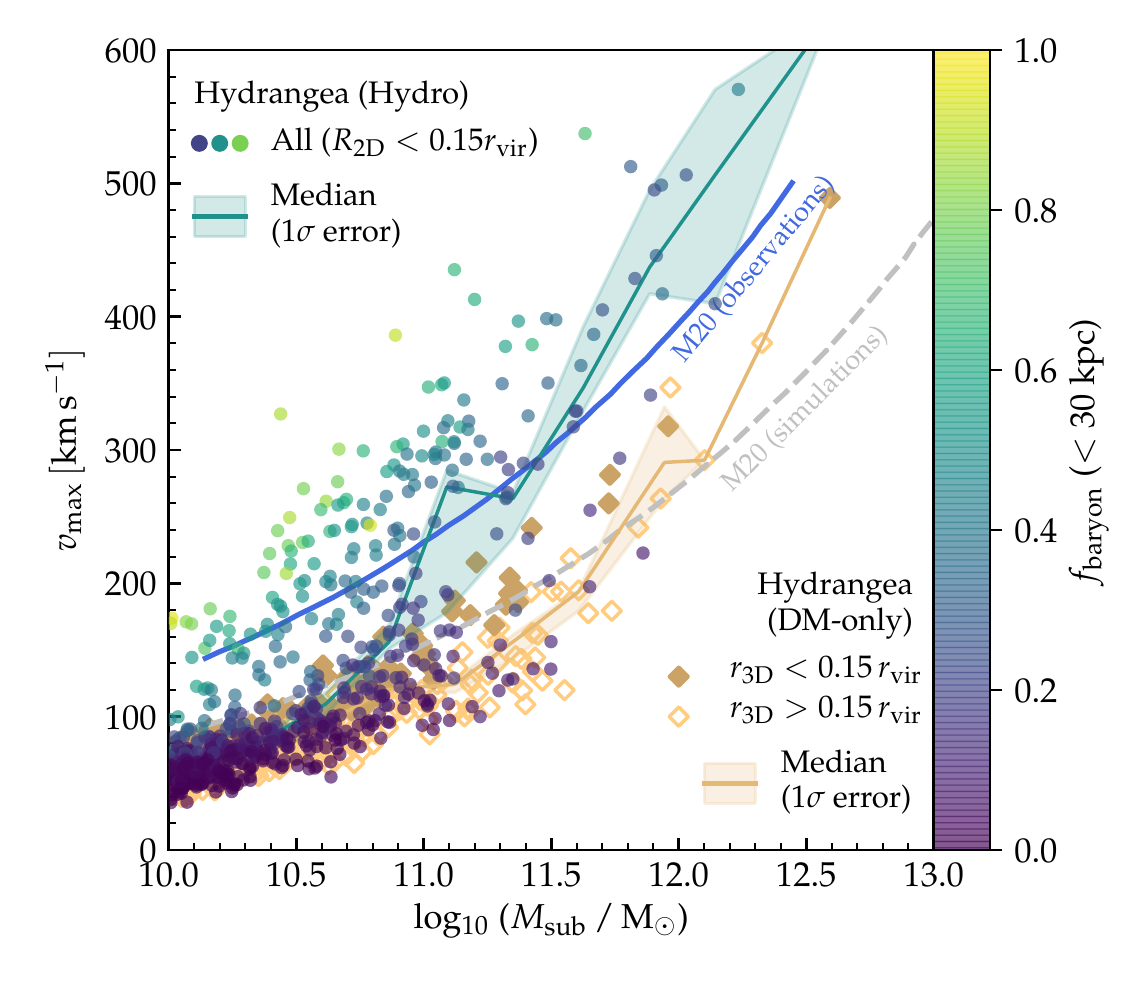}
	\caption{As Fig.~\ref{fig:vmax_msub}, but comparing simulations with and without baryons. DM-only simulations are shown as orange diamonds, filled (open) for subhaloes within (outside) a 3D radius of $0.15\, \rvir$. The hydrodynamic simulations are represented by circles, coloured according to the subhalo baryon fraction within 30 proper kpc. Running medians and their $1\sigma$ uncertainties are shown as solid lines and shaded bands, respectively. Baryons cause a $\approx$\,2-fold increase in $\vmax$ at fixed mass, with higher $\vmax$ corresponding to a higher baryon fraction.}	
	\label{fig:vmax_msub_dm}
\end{figure}

To examine the impact of baryons on the properties of cluster subhaloes, we show in Fig.~\ref{fig:vmax_msub_dm} the $\msub$--$\vmax$ relation predicted by the DM-only analogue of the Hydrangea simulation suite (i.e. evolving the same initial conditions with gravity only). These are shown as brown filled (orange open) diamonds for subhaloes within (outside) a 3D radius of 0.15 $\rvir$ from the cluster centre. For reference, the prediction from the hydrodynamical simulation, including baryons, is shown as filled circles. We here use the properties from the \textsc{subfind} catalogue for both simulation suites, because the Cantor re-processing (see Section \ref{sec:subhaloes}) has not yet been completed for the dark-matter only simulations.

It is evident that the $\msub$--$\vmax$ relation is offset to higher velocities by the inclusion of baryons. This difference is clearest at the massive end ($\msub > 10^{11}\,\msun$) -- where the DM-only simulations predict a relation $\approx$50 per cent below that inferred observationally by \citetalias{Meneghetti_et_al_2020}, with noticeably less difference between those subhaloes within and outside of 0.15 $\rvir$ -- but even at $\sim$10$^{10}\,\msun$, only the hydrodynamic simulations exhibit a tail of subhaloes extending to $\vmax > 100$ km/s.

A complementary diagnostic for the role of baryons in the $\msub$--$\vmax$ relation is the baryon mass fraction of subhaloes in the hydrodynamical simulation, as indicated by the colour of the filled circles in Fig.~\ref{fig:vmax_msub_dm}. While isolated galaxies are dominated by dark matter except possibly in the central few kpc (see e.g.~fig.~6 of \citealt{Schaller_et_al_2015a}), many of the cluster subhaloes reach baryon fractions $\gtrsim$50 per cent within 30 kpc, with a clear correlation between $f_\mathrm{baryon}$ and $\vmax$. As shown by \citet{Armitage_et_al_2019}, these high baryon fractions are the result of preferential stripping of dark matter from galaxies, while their stellar mass remains largely intact (see also \citealt{Bahe_et_al_2019} and \citealt{Joshi_et_al_2019}). The compact, high $\vmax$ subhaloes predicted by Hydrangea are therefore most likely the result of their well-resolved, centrally concentrated stellar components being able to withstand tidal stripping in a realistic way\footnote{Consistent with this interpretation, Fig.~\ref{fig:vmax_msub_dm} shows that subhaloes in the hydrodynamic simulations tend to be more massive than in the DM-only runs. The dominant difference is however that in $\vmax$: subhaloes with $\msub > 10^{11.5}\,\msun$ in the hydrodynamic simulations, for example, have on average a 32 per cent higher $\msub$ but 78 per cent higher $\vmax$ than their particle-matched subhaloes in the DM-only runs.}.

\subsection{Comparison to Illustris-TNG300}
We have shown that Hydrangea is an example of a $\Lambda$CDM simulation that predicts subhaloes with as high $\vmax$ as inferred from strong lensing observations, but the key role of baryons leaves the possibility that this is merely a fortuitous coincidence. To test this, we have repeated our analysis with the highest-resolution Illustris-TNG300 simulation \citep{Marinacci_et_al_2018, Naiman_et_al_2018, Nelson_et_al_2018,  Pillepich_et_al_2018, Springel_et_al_2018}, the only other simulation that models massive galaxy clusters at comparable resolution (a factor of six higher baryon mass than Hydrangea). Illustris-TNG300 contains sub-grid models for the same astrophysical processes as Hydrangea, but implemented in (mostly) different ways; in addition it is based on the fundamentally different hydrodynamics code \textsc{arepo} \citep{Springel_2010} that also includes magnetic fields \citep{Pakmor_Springel_2013}.

\begin{figure}
	\includegraphics[width=\columnwidth]{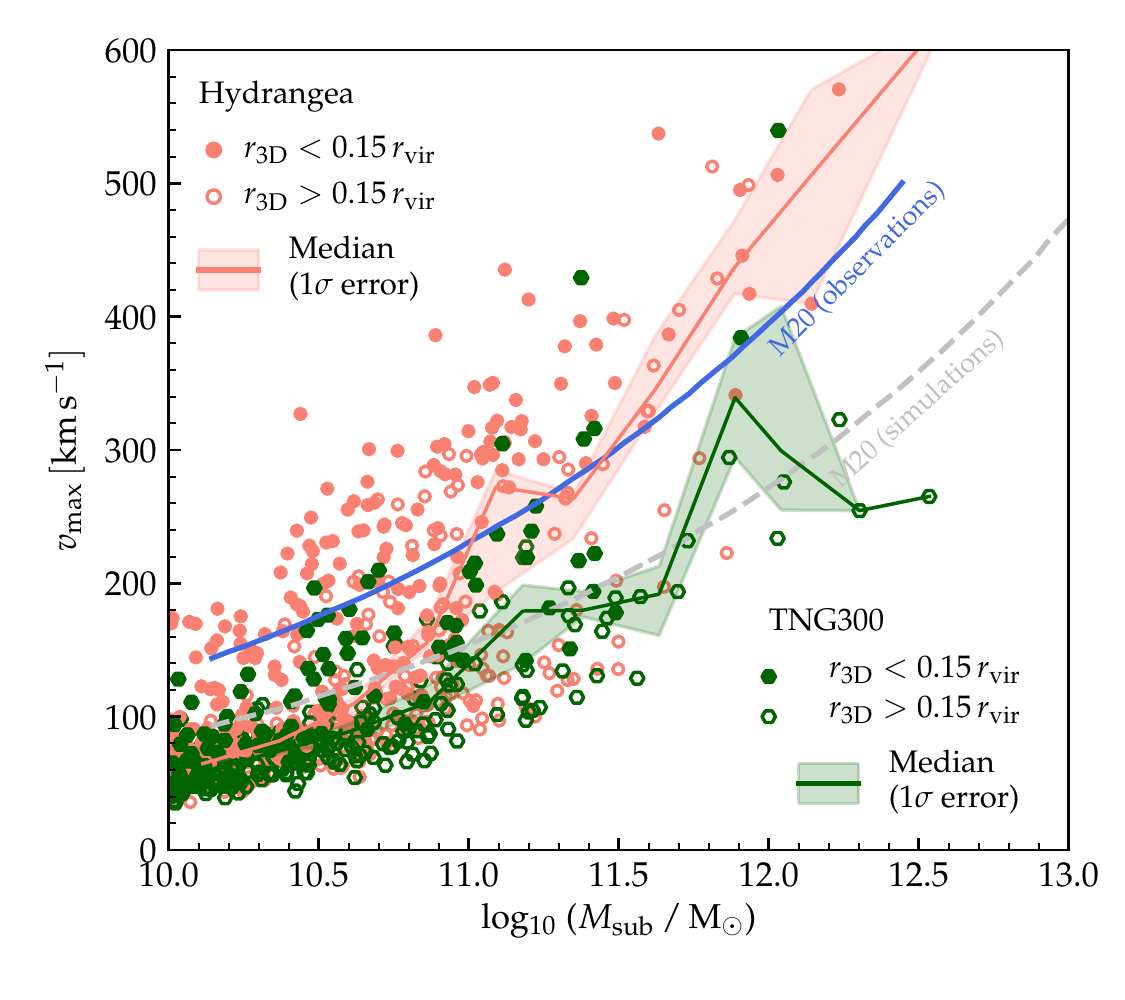}
	\caption{Comparison between Hydrangea and TNG300 in the $M_\mathrm{sub}$--$\vmax$ plane. Light red circles represent subhaloes from Hydrangea, dark green hexagons those from TNG300; in both cases, filled (open) symbols show subhaloes within (outside) a 3D radius of $0.15 \rvir$ and running medians with their 1$\sigma$ uncertainties are indicated as solid lines and light shaded bands. As in Fig.~\ref{fig:vmax_msub}, the blue solid and grey dashed lines trace the observed and simulated relation from \citetalias{Meneghetti_et_al_2020}. Although the overall population in TNG300 lies consistently below the observed relation, both simulations agree for massive subhaloes that are physically close to the cluster centre.}
	\label{fig:vmax_msub_tng}
\end{figure}

From the publicly available data \citep{Nelson_et_al_2019}, we obtain $\msub$ and $\vmax$ of subhaloes in 9 clusters with $\mvir > 5 \times 10^{14}\,\msun$ at redshift $z = 0.42$ (snapshot 71), a total of 342 within $r_\mathrm{2D} < 0.15\, \rvir$ (projected in the simulation \emph{xy} plane). These are compared to Hydrangea -- directly form \textsc{subfind} in both cases -- in Fig.~\ref{fig:vmax_msub_tng}, again distinguishing subhaloes within and outside of a 3D radius of 0.15 $\rvir$ from the cluster centre (filled and open symbols, respectively) with combined medians and their 1$\sigma$ uncertainties shown by the correspondingly coloured solid lines and shaded bands.
	
In general, the TNG300 subhaloes (dark green) show a bias towards lower $\vmax$ values at fixed $\msub$ compared to Hydrangea (light red), although this is far from uniform. At the lowest subhalo masses we probe ($\msub \approx 10^{10}\,\msun$), the ``main sequence'' of subhaloes agrees remarkably well between both simulations ($\vmax \approx 60$ km/s), while subhaloes with $\msub = 10^{12}\,\msun$ outside of 0.15 $\rvir$ (open symbols) have $\vmax$ of almost a factor of 2 lower than their Hydrangea counterparts. One possible explanation is that these ``outer'' subhaloes tend to lie at larger 3D radii in TNG300 (median $r_\mathrm{3D} = 0.55\,\rvir$) than in Hydrangea (0.33 $\rvir$). We note in particular, though, that the central massive subhaloes (filled dark green circles in the right half of the plot) agree well with their Hydrangea analogues: both predict similar high $\vmax$ for subhaloes near the centre of massive clusters.

\section{Summary and conclusions}
\label{sec:summary}
Motivated by results in the recent literature \citep[M20]{Meneghetti_et_al_2020} that massive galaxy clusters contain more concentrated substructures than predicted by $\Lambda$CDM simulations, we have compared these observations to massive galaxy clusters with $\mvir > 5\times 10^{14}\,\msun$ at $z \approx 0.4$ in the Hydrangea simulation suite \citep{Bahe_et_al_2017,Barnes_et_al_2017}. We find the following:

\begin{enumerate}
\item The subhalo mass function in the (projected) centre of galaxy clusters is realistically predicted by simulations (Fig.~\ref{fig:mass_function}).
\item Simulated subhaloes show a factor $\approx$2 scatter in maximum circular velocity $\vmax$ at fixed mass, with a general trend that agrees well with the observational inference (Fig.~\ref{fig:vmax_msub}) .
\item At fixed mass, subhaloes in simulations with baryons have up to two times higher $\vmax$ than predicted by gravity-only simulations; the offset is larger for subhaloes with a higher baryon fraction. This strongly suggests that dense stellar cores that can resist tidal stripping are key in explaining the observed strong lensing signals (Fig.~\ref{fig:vmax_msub_dm}).
\item The Illustris-TNG300 simulation predicts similarly high $\vmax$ for central cluster subhaloes as Hydrangea (Fig.~\ref{fig:vmax_msub_tng}). The ability to form subhaloes consistent with strong lensing data is therefore a common success of modern, high-resolution cosmological hydrodynamic simulations.
\end{enumerate}

In conclusion, the Hydrangea simulations demonstrate that the observed abundance and concentration of cluster substructures are compatible with $\Lambda$CDM predictions, with no evidence of a fundamental tension. Instead, the impact of baryons on cluster substructures -- also recently highlighted by \citet{Haggar_et_al_2021} -- suggests that strong lensing events in clusters can provide attractive calibration constraints for future simulations.

\section*{Acknowledgements}
I thank the anonymous referee for helpful comments that improved the presentation of results, Andrew Robertson and Joop Schaye for helpful discussions, and Massimo Meneghetti for helpful comments on an earlier version of this manuscript. 

YMB gratefully acknowledges funding from the Netherlands Organization for Scientific Research (NWO) through Veni grant number 639.041.751. The Hydrangea simulations were in part performed on the German federal maximum performance computer ``HazelHen'' at HLRS Stuttgart, under project GCS-HYDA / ID 44067, through the Gauss Center for Supercomputing project ``Hydrangea''. This work used the DiRAC@Durham facility managed by the Institute for Computational Cosmology on behalf of the STFC DiRAC HPC Facility (www.dirac.ac.uk). The equipment was funded by BEIS capital funding via STFC capital grants ST/K00042X/1, ST/P002293/1, ST/R002371/1 and ST/S002502/1, Durham University and STFC operations grant ST/R000832/1. DiRAC is part of the National e-Infrastructure.

\section*{Data Availability}

The Hydrangea simulation data are available from the author, ahead of their impending public release. IllustrisTNG-300 data are publicly available at \texttt{www.tng-project.org} \citep{Nelson_et_al_2019}. All plot data are available at \url{https://home.strw.leidenuniv.nl/~bahe/papers.html}.



\bibliographystyle{mnras}
\bibliography{lensing} 

\begin{thebibliography}{}
\makeatletter
\relax
\def\mn@urlcharsother{\let\do\@makeother \do\$\do\&\do\#\do\^\do\_\do\%\do\~}
\def\mn@doi{\begingroup\mn@urlcharsother \@ifnextchar [ {\mn@doi@}
  {\mn@doi@[]}}
\def\mn@doi@[#1]#2{\def\@tempa{#1}\ifx\@tempa\@empty \href
  {http://dx.doi.org/#2} {doi:#2}\else \href {http://dx.doi.org/#2} {#1}\fi
  \endgroup}
\def\mn@eprint#1#2{\mn@eprint@#1:#2::\@nil}
\def\mn@eprint@arXiv#1{\href {http://arxiv.org/abs/#1} {{\tt arXiv:#1}}}
\def\mn@eprint@dblp#1{\href {http://dblp.uni-trier.de/rec/bibtex/#1.xml}
  {dblp:#1}}
\def\mn@eprint@#1:#2:#3:#4\@nil{\def\@tempa {#1}\def\@tempb {#2}\def\@tempc
  {#3}\ifx \@tempc \@empty \let \@tempc \@tempb \let \@tempb \@tempa \fi \ifx
  \@tempb \@empty \def\@tempb {arXiv}\fi \@ifundefined
  {mn@eprint@\@tempb}{\@tempb:\@tempc}{\expandafter \expandafter \csname
  mn@eprint@\@tempb\endcsname \expandafter{\@tempc}}}

\bibitem[\protect\citeauthoryear{{Ahad}, {Bah{\'e}}, {Hoekstra}, {van der Burg}
   \& {Muzzin}}{{Ahad} et~al.}{2020}]{Ahad_et_al_2021}
{Ahad} L.~S.,  {Bah{\'e}} Y.~M.,  {Hoekstra} H.,  {van der Burg} R. F.~J.,
  {Muzzin} A.,  2020, arXiv e-prints, \href
  {https://ui.adsabs.harvard.edu/abs/2020arXiv201016195L} {p. arXiv:2010.16195}

\bibitem[\protect\citeauthoryear{{Armitage}, {Kay}, {Barnes}, {Bah{\'e}}  \&
  {Dalla Vecchia}}{{Armitage} et~al.}{2019}]{Armitage_et_al_2019}
{Armitage} T.~J.,  {Kay} S.~T.,  {Barnes} D.~J.,  {Bah{\'e}} Y.~M.,   {Dalla
  Vecchia} C.,  2019, \mn@doi [\mnras] {10.1093/mnras/sty2921}, \href
  {https://ui.adsabs.harvard.edu/abs/2019MNRAS.482.3308A} {482, 3308}

\bibitem[\protect\citeauthoryear{{Bah{\'e}} et~al.,}{{Bah{\'e}}
  et~al.}{2017}]{Bahe_et_al_2017}
{Bah{\'e}} Y.~M.,  et~al., 2017, \mn@doi [\mnras] {10.1093/mnras/stx1403},
  \href {https://ui.adsabs.harvard.edu/abs/2017MNRAS.470.4186B} {470, 4186}

\bibitem[\protect\citeauthoryear{{Bah{\'e}} et~al.,}{{Bah{\'e}}
  et~al.}{2019}]{Bahe_et_al_2019}
{Bah{\'e}} Y.~M.,  et~al., 2019, \mn@doi [\mnras] {10.1093/mnras/stz361}, \href
  {https://ui.adsabs.harvard.edu/abs/2019MNRAS.485.2287B} {485, 2287}

\bibitem[\protect\citeauthoryear{{Balestra} et~al.,}{{Balestra}
  et~al.}{2016}]{Balestra_et_al_2016}
{Balestra} I.,  et~al., 2016, \mn@doi [\apjs] {10.3847/0067-0049/224/2/33},
  \href {https://ui.adsabs.harvard.edu/abs/2016ApJS..224...33B} {224, 33}

\bibitem[\protect\citeauthoryear{{Barnes} et~al.,}{{Barnes}
  et~al.}{2017}]{Barnes_et_al_2017}
{Barnes} D.~J.,  et~al., 2017, \mn@doi [\mnras] {10.1093/mnras/stx1647}, \href
  {https://ui.adsabs.harvard.edu/abs/2017MNRAS.471.1088B} {471, 1088}

\bibitem[\protect\citeauthoryear{{Behroozi}, {Wechsler}  \& {Wu}}{{Behroozi}
  et~al.}{2013}]{Behroozi_et_al_2013}
{Behroozi} P.~S.,  {Wechsler} R.~H.,   {Wu} H.-Y.,  2013, \mn@doi [\apj]
  {10.1088/0004-637X/762/2/109}, \href
  {https://ui.adsabs.harvard.edu/abs/2013ApJ...762..109B} {762, 109}

\bibitem[\protect\citeauthoryear{{Bergamini} et~al.,}{{Bergamini}
  et~al.}{2019}]{Bergamini_et_al_2019}
{Bergamini} P.,  et~al., 2019, \mn@doi [\aap] {10.1051/0004-6361/201935974},
  \href {https://ui.adsabs.harvard.edu/abs/2019A&A...631A.130B} {631, A130}

\bibitem[\protect\citeauthoryear{{Biviano} et~al.,}{{Biviano}
  et~al.}{2013}]{Biviano_et_al_2013}
{Biviano} A.,  et~al., 2013, \mn@doi [\aap] {10.1051/0004-6361/201321955},
  \href {https://ui.adsabs.harvard.edu/abs/2013A&A...558A...1B} {558, A1}

\bibitem[\protect\citeauthoryear{{Blumenthal}, {Faber}, {Primack}  \&
  {Rees}}{{Blumenthal} et~al.}{1984}]{Blumenthal_et_al_1984}
{Blumenthal} G.~R.,  {Faber} S.~M.,  {Primack} J.~R.,   {Rees} M.~J.,  1984,
  \mn@doi [\nat] {10.1038/311517a0}, \href
  {https://ui.adsabs.harvard.edu/abs/1984Natur.311..517B} {311, 517}

\bibitem[\protect\citeauthoryear{{Bryan} \& {Norman}}{{Bryan} \&
  {Norman}}{1998}]{Bryan_Norman_1998}
{Bryan} G.~L.,  {Norman} M.~L.,  1998, \mn@doi [\apj] {10.1086/305262}, \href
  {https://ui.adsabs.harvard.edu/abs/1998ApJ...495...80B} {495, 80}

\bibitem[\protect\citeauthoryear{{Caminha} et~al.,}{{Caminha}
  et~al.}{2016}]{Caminha_et_al_2016}
{Caminha} G.~B.,  et~al., 2016, \mn@doi [\aap] {10.1051/0004-6361/201527670},
  \href {https://ui.adsabs.harvard.edu/abs/2016A&A...587A..80C} {587, A80}

\bibitem[\protect\citeauthoryear{{Crain} et~al.,}{{Crain}
  et~al.}{2015}]{Crain_et_al_2015}
{Crain} R.~A.,  et~al., 2015, \mn@doi [\mnras] {10.1093/mnras/stv725}, \href
  {https://ui.adsabs.harvard.edu/abs/2015MNRAS.450.1937C} {450, 1937}

\bibitem[\protect\citeauthoryear{{Cui} et~al.,}{{Cui}
  et~al.}{2018}]{Cui_et_al_2018}
{Cui} W.,  et~al., 2018, \mn@doi [\mnras] {10.1093/mnras/sty2111}, \href
  {https://ui.adsabs.harvard.edu/abs/2018MNRAS.480.2898C} {480, 2898}

\bibitem[\protect\citeauthoryear{{Dalla Vecchia} \& {Schaye}}{{Dalla Vecchia}
  \& {Schaye}}{2012}]{DallaVecchia_Schaye_2012}
{Dalla Vecchia} C.,  {Schaye} J.,  2012, \mn@doi [\mnras]
  {10.1111/j.1365-2966.2012.21704.x}, \href
  {https://ui.adsabs.harvard.edu/abs/2012MNRAS.426..140D} {426, 140}

\bibitem[\protect\citeauthoryear{{Despali} \& {Vegetti}}{{Despali} \&
  {Vegetti}}{2017}]{Despali_Vegetti_2017}
{Despali} G.,  {Vegetti} S.,  2017, \mn@doi [\mnras] {10.1093/mnras/stx966},
  \href {https://ui.adsabs.harvard.edu/abs/2017MNRAS.469.1997D} {469, 1997}

\bibitem[\protect\citeauthoryear{{Dolag}, {Borgani}, {Murante}  \&
  {Springel}}{{Dolag} et~al.}{2009}]{Dolag_et_al_2009}
{Dolag} K.,  {Borgani} S.,  {Murante} G.,   {Springel} V.,  2009, \mn@doi
  [\mnras] {10.1111/j.1365-2966.2009.15034.x}, \href
  {https://ui.adsabs.harvard.edu/abs/2009MNRAS.399..497D} {399, 497}

\bibitem[\protect\citeauthoryear{{Duffy}, {Schaye}, {Kay}, {Dalla Vecchia},
  {Battye}  \& {Booth}}{{Duffy} et~al.}{2010}]{Duffy_et_al_2010}
{Duffy} A.~R.,  {Schaye} J.,  {Kay} S.~T.,  {Dalla Vecchia} C.,  {Battye}
  R.~A.,   {Booth} C.~M.,  2010, \mn@doi [\mnras]
  {10.1111/j.1365-2966.2010.16613.x}, \href
  {https://ui.adsabs.harvard.edu/abs/2010MNRAS.405.2161D} {405, 2161}

\bibitem[\protect\citeauthoryear{{Gao}, {Navarro}, {Frenk}, {Jenkins},
  {Springel}  \& {White}}{{Gao} et~al.}{2012}]{Gao_et_al_2012}
{Gao} L.,  {Navarro} J.~F.,  {Frenk} C.~S.,  {Jenkins} A.,  {Springel} V.,
  {White} S.~D.~M.,  2012, \mn@doi [\mnras] {10.1111/j.1365-2966.2012.21564.x},
  \href {https://ui.adsabs.harvard.edu/abs/2012MNRAS.425.2169G} {425, 2169}

\bibitem[\protect\citeauthoryear{{Grillo} et~al.,}{{Grillo}
  et~al.}{2015}]{Grillo_et_al_2015}
{Grillo} C.,  et~al., 2015, \mn@doi [\apj] {10.1088/0004-637X/800/1/38}, \href
  {https://ui.adsabs.harvard.edu/abs/2015ApJ...800...38G} {800, 38}

\bibitem[\protect\citeauthoryear{{Haggar}, {Pearce}, {Gray}, {Knebe}  \&
  {Yepes}}{{Haggar} et~al.}{2021}]{Haggar_et_al_2021}
{Haggar} R.,  {Pearce} F.~R.,  {Gray} M.~E.,  {Knebe} A.,   {Yepes} G.,  2021,
  \mn@doi [\mnras] {10.1093/mnras/stab064}, \href
  {https://ui.adsabs.harvard.edu/abs/2021MNRAS.tmp...89H} {}

\bibitem[\protect\citeauthoryear{{Helmi}}{{Helmi}}{2008}]{Helmi_2008}
{Helmi} A.,  2008, \mn@doi [\aapr] {10.1007/s00159-008-0009-6}, \href
  {https://ui.adsabs.harvard.edu/abs/2008A&ARv..15..145H} {15, 145}

\bibitem[\protect\citeauthoryear{{Joshi}, {Parker}, {Wadsley}  \&
  {Keller}}{{Joshi} et~al.}{2019}]{Joshi_et_al_2019}
{Joshi} G.~D.,  {Parker} L.~C.,  {Wadsley} J.,   {Keller} B.~W.,  2019, \mn@doi
  [\mnras] {10.1093/mnras/sty3119}, \href
  {https://ui.adsabs.harvard.edu/abs/2019MNRAS.483..235J} {483, 235}

\bibitem[\protect\citeauthoryear{{Jullo}, {Kneib}, {Limousin},
  {El{\'\i}asd{\'o}ttir}, {Marshall}  \& {Verdugo}}{{Jullo}
  et~al.}{2007}]{Jullo_et_al_2007}
{Jullo} E.,  {Kneib} J.~P.,  {Limousin} M.,  {El{\'\i}asd{\'o}ttir} {\'A}.,
  {Marshall} P.~J.,   {Verdugo} T.,  2007, \mn@doi [New Journal of Physics]
  {10.1088/1367-2630/9/12/447}, \href
  {https://ui.adsabs.harvard.edu/abs/2007NJPh....9..447J} {9, 447}

\bibitem[\protect\citeauthoryear{{Kneib} \& {Natarajan}}{{Kneib} \&
  {Natarajan}}{2011}]{Kneib_Natarajan_2011}
{Kneib} J.-P.,  {Natarajan} P.,  2011, \mn@doi [\aapr]
  {10.1007/s00159-011-0047-3}, \href
  {https://ui.adsabs.harvard.edu/abs/2011A&ARv..19...47K} {19, 47}

\bibitem[\protect\citeauthoryear{{Lotz} et~al.,}{{Lotz}
  et~al.}{2017}]{Lotz_et_al_2017}
{Lotz} J.~M.,  et~al., 2017, \mn@doi [\apj] {10.3847/1538-4357/837/1/97}, \href
  {https://ui.adsabs.harvard.edu/abs/2017ApJ...837...97L} {837, 97}

\bibitem[\protect\citeauthoryear{{Marinacci} et~al.,}{{Marinacci}
  et~al.}{2018}]{Marinacci_et_al_2018}
{Marinacci} F.,  et~al., 2018, \mn@doi [\mnras] {10.1093/mnras/sty2206}, \href
  {https://ui.adsabs.harvard.edu/abs/2018MNRAS.480.5113M} {480, 5113}

\bibitem[\protect\citeauthoryear{{McAlpine} et~al.,}{{McAlpine}
  et~al.}{2016}]{McAlpine_et_al_2016}
{McAlpine} S.,  et~al., 2016, \mn@doi [Astronomy and Computing]
  {10.1016/j.ascom.2016.02.004}, \href
  {https://ui.adsabs.harvard.edu/abs/2016A&C....15...72M} {15, 72}

\bibitem[\protect\citeauthoryear{{Meneghetti} et~al.,}{{Meneghetti}
  et~al.}{2020}]{Meneghetti_et_al_2020}
{Meneghetti} M.,  et~al., 2020, \mn@doi [Science] {10.1126/science.aax5164},
  \href {https://ui.adsabs.harvard.edu/abs/2020Sci...369.1347M} {369, 1347}

\bibitem[\protect\citeauthoryear{{Muldrew}, {Pearce}  \& {Power}}{{Muldrew}
  et~al.}{2011}]{Muldrew_et_al_2011}
{Muldrew} S.~I.,  {Pearce} F.~R.,   {Power} C.,  2011, \mn@doi [\mnras]
  {10.1111/j.1365-2966.2010.17636.x}, \href
  {http://adsabs.harvard.edu/abs/2011MNRAS.410.2617M} {410, 2617}

\bibitem[\protect\citeauthoryear{{Naiman} et~al.,}{{Naiman}
  et~al.}{2018}]{Naiman_et_al_2018}
{Naiman} J.~P.,  et~al., 2018, \mn@doi [\mnras] {10.1093/mnras/sty618}, \href
  {https://ui.adsabs.harvard.edu/abs/2018MNRAS.477.1206N} {477, 1206}

\bibitem[\protect\citeauthoryear{{Nelson} et~al.,}{{Nelson}
  et~al.}{2018}]{Nelson_et_al_2018}
{Nelson} D.,  et~al., 2018, \mn@doi [\mnras] {10.1093/mnras/stx3040}, \href
  {https://ui.adsabs.harvard.edu/abs/2018MNRAS.475..624N} {475, 624}

\bibitem[\protect\citeauthoryear{{Nelson} et~al.,}{{Nelson}
  et~al.}{2019}]{Nelson_et_al_2019}
{Nelson} D.,  et~al., 2019, \mn@doi [Computational Astrophysics and Cosmology]
  {10.1186/s40668-019-0028-x}, \href
  {https://ui.adsabs.harvard.edu/abs/2019ComAC...6....2N} {6, 2}

\bibitem[\protect\citeauthoryear{{Pakmor} \& {Springel}}{{Pakmor} \&
  {Springel}}{2013}]{Pakmor_Springel_2013}
{Pakmor} R.,  {Springel} V.,  2013, \mn@doi [\mnras] {10.1093/mnras/stt428},
  \href {https://ui.adsabs.harvard.edu/abs/2013MNRAS.432..176P} {432, 176}

\bibitem[\protect\citeauthoryear{{Pearce}, {Kay}, {Barnes}, {Bahe}  \&
  {Bower}}{{Pearce} et~al.}{2020}]{Pearce_et_al_2020}
{Pearce} F.~A.,  {Kay} S.~T.,  {Barnes} D.~J.,  {Bahe} Y.~M.,   {Bower} R.~G.,
  2020, arXiv e-prints, \href
  {https://ui.adsabs.harvard.edu/abs/2020arXiv200512391P} {p. arXiv:2005.12391}

\bibitem[\protect\citeauthoryear{{Pillepich} et~al.,}{{Pillepich}
  et~al.}{2018}]{Pillepich_et_al_2018}
{Pillepich} A.,  et~al., 2018, \mn@doi [\mnras] {10.1093/mnras/stx3112}, \href
  {https://ui.adsabs.harvard.edu/abs/2018MNRAS.475..648P} {475, 648}

\bibitem[\protect\citeauthoryear{{Planelles}, {Borgani}, {Fabjan}, {Killedar},
  {Murante}, {Granato}, {Ragone-Figueroa}  \& {Dolag}}{{Planelles}
  et~al.}{2014}]{Planelles_et_al_2014}
{Planelles} S.,  {Borgani} S.,  {Fabjan} D.,  {Killedar} M.,  {Murante} G.,
  {Granato} G.~L.,  {Ragone-Figueroa} C.,   {Dolag} K.,  2014, \mn@doi [\mnras]
  {10.1093/mnras/stt2141}, \href
  {https://ui.adsabs.harvard.edu/abs/2014MNRAS.438..195P} {438, 195}

\bibitem[\protect\citeauthoryear{{Postman} et~al.,}{{Postman}
  et~al.}{2012}]{Postman_et_al_2012}
{Postman} M.,  et~al., 2012, \mn@doi [\apjs] {10.1088/0067-0049/199/2/25},
  \href {https://ui.adsabs.harvard.edu/abs/2012ApJS..199...25P} {199, 25}

\bibitem[\protect\citeauthoryear{{Robertson}}{{Robertson}}{2021}]{Robertson_2021}
{Robertson} A.,  2021, arXiv e-prints, \href
  {https://ui.adsabs.harvard.edu/abs/2021arXiv210112067R} {p. arXiv:2101.12067}

\bibitem[\protect\citeauthoryear{{Rosas-Guevara} et~al.,}{{Rosas-Guevara}
  et~al.}{2015}]{Rosas-Guevara_et_al_2015}
{Rosas-Guevara} Y.~M.,  et~al., 2015, \mn@doi [\mnras] {10.1093/mnras/stv2056},
  \href {https://ui.adsabs.harvard.edu/abs/2015MNRAS.454.1038R} {454, 1038}

\bibitem[\protect\citeauthoryear{{Sartoris} et~al.,}{{Sartoris}
  et~al.}{2020}]{Sartoris_et_al_2020}
{Sartoris} B.,  et~al., 2020, \mn@doi [\aap] {10.1051/0004-6361/202037521},
  \href {https://ui.adsabs.harvard.edu/abs/2020A&A...637A..34S} {637, A34}

\bibitem[\protect\citeauthoryear{{Schaller} et~al.,}{{Schaller}
  et~al.}{2015a}]{Schaller_et_al_2015a}
{Schaller} M.,  et~al., 2015a, \mn@doi [\mnras] {10.1093/mnras/stv1067}, \href
  {https://ui.adsabs.harvard.edu/abs/2015MNRAS.451.1247S} {451, 1247}

\bibitem[\protect\citeauthoryear{{Schaller}, {Dalla Vecchia}, {Schaye},
  {Bower}, {Theuns}, {Crain}, {Furlong}  \& {McCarthy}}{{Schaller}
  et~al.}{2015b}]{Schaller_et_al_2015}
{Schaller} M.,  {Dalla Vecchia} C.,  {Schaye} J.,  {Bower} R.~G.,  {Theuns} T.,
   {Crain} R.~A.,  {Furlong} M.,   {McCarthy} I.~G.,  2015b, \mn@doi [\mnras]
  {10.1093/mnras/stv2169}, \href
  {https://ui.adsabs.harvard.edu/abs/2015MNRAS.454.2277S} {454, 2277}

\bibitem[\protect\citeauthoryear{{Schaye}}{{Schaye}}{2004}]{Schaye_2004}
{Schaye} J.,  2004, \mn@doi [\apj] {10.1086/421232}, \href
  {https://ui.adsabs.harvard.edu/abs/2004ApJ...609..667S} {609, 667}

\bibitem[\protect\citeauthoryear{{Schaye} \& {Dalla Vecchia}}{{Schaye} \&
  {Dalla Vecchia}}{2008}]{Schaye_DallaVecchia_2008}
{Schaye} J.,  {Dalla Vecchia} C.,  2008, \mn@doi [\mnras]
  {10.1111/j.1365-2966.2007.12639.x}, \href
  {https://ui.adsabs.harvard.edu/abs/2008MNRAS.383.1210S} {383, 1210}

\bibitem[\protect\citeauthoryear{{Schaye} et~al.,}{{Schaye}
  et~al.}{2015}]{Schaye_et_al_2015}
{Schaye} J.,  et~al., 2015, \mn@doi [\mnras] {10.1093/mnras/stu2058}, \href
  {https://ui.adsabs.harvard.edu/abs/2015MNRAS.446..521S} {446, 521}

\bibitem[\protect\citeauthoryear{{Shipp} et~al.,}{{Shipp}
  et~al.}{2018}]{Shipp_et_al_2018}
{Shipp} N.,  et~al., 2018, \mn@doi [\apj] {10.3847/1538-4357/aacdab}, \href
  {https://ui.adsabs.harvard.edu/abs/2018ApJ...862..114S} {862, 114}

\bibitem[\protect\citeauthoryear{{Springel}}{{Springel}}{2010}]{Springel_2010}
{Springel} V.,  2010, \mn@doi [\mnras] {10.1111/j.1365-2966.2009.15715.x},
  \href {https://ui.adsabs.harvard.edu/abs/2010MNRAS.401..791S} {401, 791}

\bibitem[\protect\citeauthoryear{{Springel}, {White}, {Tormen}  \&
  {Kauffmann}}{{Springel} et~al.}{2001}]{Springel_et_al_2001}
{Springel} V.,  {White} S.~D.~M.,  {Tormen} G.,   {Kauffmann} G.,  2001,
  \mn@doi [\mnras] {10.1046/j.1365-8711.2001.04912.x}, \href
  {http://adsabs.harvard.edu/abs/2001MNRAS.328..726S} {328, 726}

\bibitem[\protect\citeauthoryear{{Springel} et~al.,}{{Springel}
  et~al.}{2018}]{Springel_et_al_2018}
{Springel} V.,  et~al., 2018, \mn@doi [\mnras] {10.1093/mnras/stx3304}, \href
  {https://ui.adsabs.harvard.edu/abs/2018MNRAS.475..676S} {475, 676}

\bibitem[\protect\citeauthoryear{{Treu} et~al.,}{{Treu}
  et~al.}{2015}]{Treu_et_al_2015}
{Treu} T.,  et~al., 2015, \mn@doi [\apj] {10.1088/0004-637X/812/2/114}, \href
  {https://ui.adsabs.harvard.edu/abs/2015ApJ...812..114T} {812, 114}

\bibitem[\protect\citeauthoryear{{Vogelsberger}, {Marinacci}, {Torrey}  \&
  {Puchwein}}{{Vogelsberger} et~al.}{2020}]{Vogelsberger_et_al_2020}
{Vogelsberger} M.,  {Marinacci} F.,  {Torrey} P.,   {Puchwein} E.,  2020,
  \mn@doi [Nature Reviews Physics] {10.1038/s42254-019-0127-2}, \href
  {https://ui.adsabs.harvard.edu/abs/2020NatRP...2...42V} {2, 42}

\bibitem[\protect\citeauthoryear{{Wang}, {Bose}, {Frenk}, {Gao}, {Jenkins},
  {Springel}  \& {White}}{{Wang} et~al.}{2020}]{Wang_et_al_2020}
{Wang} J.,  {Bose} S.,  {Frenk} C.~S.,  {Gao} L.,  {Jenkins} A.,  {Springel}
  V.,   {White} S.~D.~M.,  2020, \mn@doi [\nat] {10.1038/s41586-020-2642-9},
  \href {https://ui.adsabs.harvard.edu/abs/2020Natur.585...39W} {585, 39}

\bibitem[\protect\citeauthoryear{{Wiersma}, {Schaye}  \& {Smith}}{{Wiersma}
  et~al.}{2009a}]{Wiersma_et_al_2009a}
{Wiersma} R. P.~C.,  {Schaye} J.,   {Smith} B.~D.,  2009a, \mn@doi [\mnras]
  {10.1111/j.1365-2966.2008.14191.x}, \href
  {https://ui.adsabs.harvard.edu/abs/2009MNRAS.393...99W} {393, 99}

\bibitem[\protect\citeauthoryear{{Wiersma}, {Schaye}, {Theuns}, {Dalla Vecchia}
   \& {Tornatore}}{{Wiersma} et~al.}{2009b}]{Wiersma_et_al_2009b}
{Wiersma} R. P.~C.,  {Schaye} J.,  {Theuns} T.,  {Dalla Vecchia} C.,
  {Tornatore} L.,  2009b, \mn@doi [\mnras] {10.1111/j.1365-2966.2009.15331.x},
  \href {https://ui.adsabs.harvard.edu/abs/2009MNRAS.399..574W} {399, 574}

\bibitem[\protect\citeauthoryear{{Yang} \& {Yu}}{{Yang} \&
  {Yu}}{2021}]{Yang_Yu_2021}
{Yang} D.,  {Yu} H.-B.,  2021, arXiv e-prints, \href
  {https://ui.adsabs.harvard.edu/abs/2021arXiv210202375Y} {p. arXiv:2102.02375}

\bibitem[\protect\citeauthoryear{{Yang}, {Mo}, {van den Bosch}, {Pasquali},
  {Li}  \& {Barden}}{{Yang} et~al.}{2007}]{Yang_et_al_2007}
{Yang} X.,  {Mo} H.~J.,  {van den Bosch} F.~C.,  {Pasquali} A.,  {Li} C.,
  {Barden} M.,  2007, \mn@doi [\apj] {10.1086/522027}, \href
  {https://ui.adsabs.harvard.edu/abs/2007ApJ...671..153Y} {671, 153}

\makeatother
\end{thebibliography}


\bsp	
\label{lastpage}
\end{document}